# Investigation of the electroplastic effect using nanoindentation

D. Andre[1], T. Burlet[1], F. Körkemeyer[2], G. Gerstein[2], S. Sandlöbes-Haut[1], S. Korte-Kerzel[1]*

[1]Institute of Physical Metallurgy and Metal Physics, RWTH Aachen University,

[2]Institute of Materials Science, Leibniz University Hannover

* korte-kerzel@imm.rwth-aachen.de (corresponding author)

## Abstract

A promising approach to deform metallic-intermetallic composite materials is the application of electric current pulses during the deformation process to achieve a lower yield strength and enhanced elongation to fracture. This is known as the electroplastic effect. We developed a novel setup to study the electroplastic effect during nanoindentation on individual phases and well-defined interfaces. Using a eutectic Al-$Al_2Cu$ alloy as a model material, we compare the electroplastic nanoindentation results to macroscopic electroplastic compression tests. The results of the micro- and macroscopic investigations reveal current induced displacement shifts and stress drops, respectively, with the first displacement shift / stress drop being higher than the subsequent ones. A higher current intensity, higher loading rate and larger pulsing interval all cause increased displacement shifts. This observation, in conjunction with the fact that the first displacement shift is highest, strongly indicates that de-pinning of dislocations from obstacles dominates the mechanical response, rather than solely thermal effects.

## 1. Introduction

The influence of electric current on the plastic deformation of metals, termed as the electroplastic effect (EPE), was first investigated by Troitskii and Likhtman [1]. The EPE was subsequently found to influence a wide range of mechanical properties, such as flow stress [2, 3], stress relaxation [4], creep [4], dislocation generation and mobility [5], brittle fracture [6], and fatigue [7]. Systematic studies were carried out to study the nature of the EPE over a variety of loading conditions and materials and the EPE was also found to affect the materials' behaviour during processing [8, 9]. Though several mechanisms for the EPE were proposed, our understanding of its nature remains at a nascent stage. Local adiabatic temperature increase due to fast joule heating was thought to be the major contributor to the EPE due to local thermal softening [10]. However, the EPE was also observed at low temperatures and high-density current pulses [1], during which the joule heating is negligible, thus indicating that a part of the EPE is intrinsically related to the interaction between drift electrons and dislocations. Early work by Troitskii et al. [1, 11] and later by Conrad et al. [5, 12] considered the effect of the 'electron wind', i.e. the drift of conduction electrons upon application of an electric potential, in assisting dislocations to overcome lattice obstacles. The 'electron wind' theory has since been widely applied and is able to predict key features of the EPE in a variety of materials and loading conditions [12]. More recently, Molotskii [13] also proposed the de-pinning of dislocations from paramagnetic obstacles by the magnetic field induced by the electric current as an alternative mechanism for the EPE in metals. Beyond the EPE, electric currents can also induce phase transformations through increased nucleation and recrystallization rates, often resulting in grain refinement and enhanced mechanical properties [14-16]. Furthermore, the effects of an electric current applied separately to any deformation or heating process has been shown to differ from those where an electric current is applied in combination with deformation [17].

Several key factors in the enhanced dislocation motion were identified, such as the effects of drift electrons on the driving force for dislocation motion, the activation energy barrier of dislocation obstacles, and the de-pinning of dislocations from paramagnetic obstacles [13]. However, until now, no method has been available which allowed a quantitative investigation of the EPE in a well-defined local environment, such as an individual grain or eutectic region, selected crystal orientations or isolated grain and phase boundaries.

Electroplastic forming, in which the EPE is exploited by passing a current through the work piece during processing, is particularly promising for materials that are otherwise brittle. An example are metallic-intermetallic composite materials that are – at least theoretically – capable of combining the high formability of metals and high strength of intermetallic phases. Plasticity in such composite alloys is predominantly carried by the formable metallic components while the brittle intermetallic phases strengthen the material. Microstructurally weak points are the interfaces between metallic and intermetallic phases due to stress concentrations promoting strain localization and failure initiation [18-23]. In general, the constituent phases and the morphology and spacing of the eutectic structures control the mechanical behaviour of eutectic composite alloys [24, 25]. Structural refinement is reported to cause an increase in strength concomitant with a loss in ductility [24, 25].

As an example of metallic-intermetallic composite materials, Al-Cu eutectic alloys exhibit low density, good castability, corrosion resistance and a wide range of available phases and microstructural morphologies. The main microstructural constituents of eutectic Al-Cu composite alloys are cubic (fcc) Al and tetragonal (bct) $Al_2Cu$ [26]. Depending on the composition, solidification and possible heat treatment conditions different volume fractions and morphologies of fcc Al and bct $Al_2Cu$ form. Additionally, some more copper-rich intermetallic phases might form as precipitates [25, 27-29].

The main drawback of Al-Cu eutectic composite alloys is brittle deformation at ambient temperatures [18, 23, 30, 31]. Early detailed characterisation of deformation microstructures of Al-Cu eutectic alloys revealed dislocation-dominated deformation in Al grains with higher dislocation densities at Al-$Al_2Cu$ interfaces while the eutectic lamellae deformed by kinking to accommodate compressive strain. Consequently, fracture was observed along $Al_2Cu$ interfaces or through $Al_2Cu$ grains [23, 31]. The Al-Cu composite is therefore a candidate material to benefit from electroplastic forming if the process enables plastic flow in the brittle phase and its interfaces, such that formability is enhanced and damage during forming reduced.

Here, we present the first local studies of the EPE at the scale of an alloy's individual microstructural components. For this, we developed a novel in-situ nanoindentation setup in which high current densities are passed through the contact during deformation. This setup therefore opens new opportunities to investigate the EPE and the underlying mechanisms. In order to allow a direct comparison across the scales, we study and compare the effects of electric pulses during deformation in a eutectic Al-Cu alloy on both, the macroscopic and microscopic scale.

# 2. Experimental procedure

## 2.1. Macroscopic electroplasticity experiments

Macroscopic electrically pulsed compression experiments were performed on bulk eutectic Al-Al$_2$Cu samples with a composition of 33.1 wt. % copper and 66.9 wt. % aluminium. The as-cast material was cut into specimens with dimensions of 2 mm x 2 mm x 4 mm (width x length x height) using electric discharge machining. Compression experiments with and without electric current pulsing during deformation were performed at a deformation rate of 0.1 mm/min using a Walter & Bai 100 universal testing machine at the Institute of Materials Science, Leibniz University Hannover. After the application of a pre-force of 80 N and an offset time to ensure proper electrical contact, electric current pulses of 7 kA, 55 V and a duration of 0.5 ms were applied every second using a high current impulse generator (further information regarding the high current impulse generator are available in [32]). The resulting current densities amounted to 1.66 ± 0.01 kA/mm$^2$ when considering the contact area prior to deformation. For comparison, compression experiments on the same material with the same sample geometry and machine settings were performed without electrical pulses. During the compression tests, the sample temperature was recorded using a FLIR Systems, ThermaCam SC3000 at a frequency of 250 Hz.

## 2.2 Microscopic electroplasticity experiments

The same eutectic Al-Al$_2$Cu alloy with a composition of 33.1. wt. % copper and 66.9 wt. % aluminium as used for the macroscopic tests, was investigated using electroplastic nanoindentation experiments (Figure 1). Furthermore, electroplastic nanoindentation was also performed on as-cast samples of the Al$_2$Cu-θ phase and the Al α-phase.

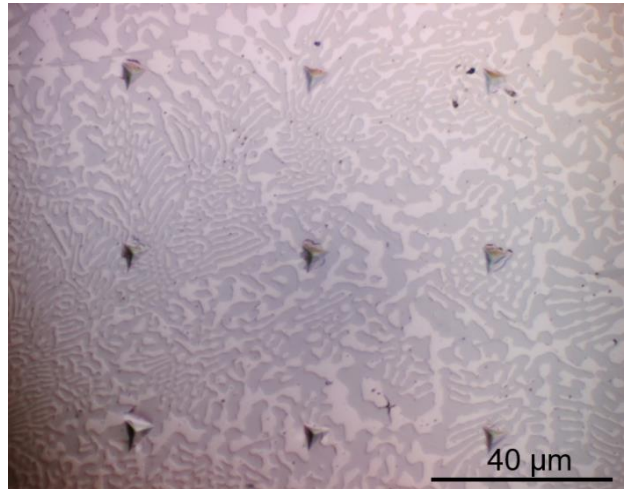

*Figure 1: Al-Al$_2$Cu eutectic alloy used for nanoindentation experiments.*

The samples were metallographically prepared and deformed using a nanoindenter (NanoTest, Micromaterials Ltd., UK) modified as part of this work (see Chapter 3.2.1. 'technical development') to allow application of electric current pulses of 100 µs in length. Using this new nanoindentation setup, four different current intensities in the range of 1.5 – 3 A and two different pulsing intervals of 1 s and 2 s using a loading and unloading rate of 20 mN/s to a final load of 500 mN were applied. Additionally, experiments with a current intensity of 2.5 A during loading and unloading with loading and unloading rates of 10 mN/s and 20 mN/s were carried out to identify the effects of electric current pulses during unloading. In order to investigate the interplay of indentation depth, current and resulting current density, additional experiments were carried out with a constant displacement rate of 20 nm/s until maximum displacements of 2000 nm and 1200 nm. To evaluate possible area functions (projected contact area as used in the Oliver and Pharr analysis [33], true area of contact between material and indenter or circumference of the plastic zone), increasing intensities of the electric current pulses scaling with each potential area function at each depth/time increment were used (see Supplementary Materials).

## 3. Results

### 3.1. Macroscopic electroplasticity experiments

### 3.1.1. Macroscopic stress-strain response

Figure 2 shows the true stress-strain curves of macroscopic compression experiments with and without electric pulses. In the beginning of the compression experiments, before electric current pulses were applied, all samples show similar stress strain responses. When applying the first electric current pulse, large stress drops of about 47.8 ± 4.8 MPa occurred (highlighted by the red box in Figure 2). The subsequently applied current pulses caused smaller stress drops of about 18.3 ± 1.6 MPa during compression. The stress-strain curves of the samples with and without electric current pulses have similar slopes. At a strain of 5% - 6% the electric current pulses were turned off to avoid electric arcing during fracture. With the shut-off of the electric current pulsing, the stress of the samples maintains a lower stress level than that of samples compressed without electric current pulses at the same strain value and fracture shortly after the final electric current pulses, see blue box in Figure 2.

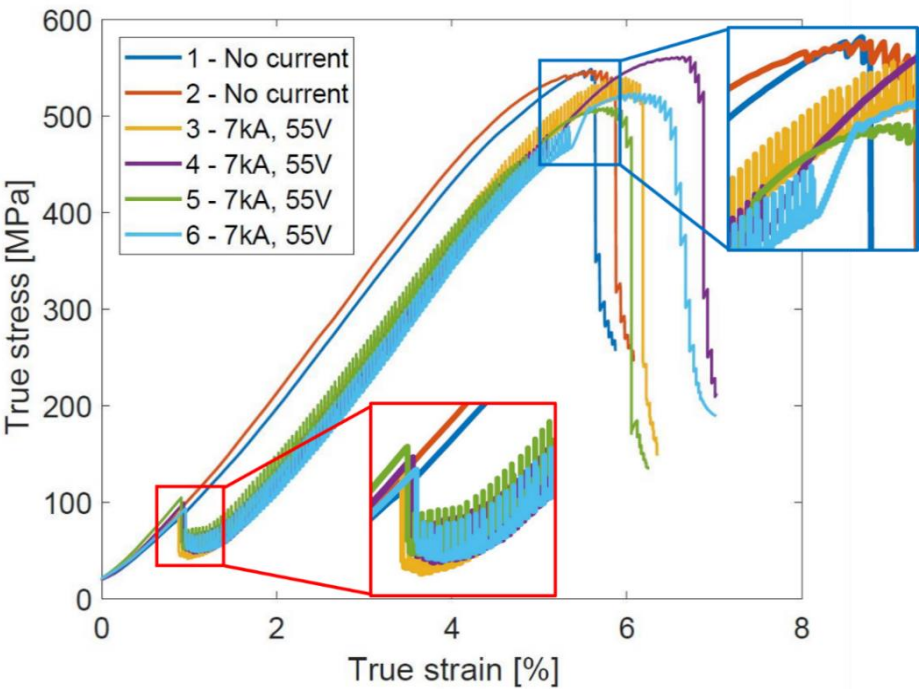

*Figure 2: True compressive stress-strain curves of samples deformed with and without the application of electric current pulses. The electric current pulses had a current of 7 kA, voltage of 55 V and duration of 0.5 ms at a frequency of 1 Hz. The initial strain rate was 0.1 mm/min.*

Generally, samples exposed to electric current pulses during compression reached higher fracture strains and exhibited lower flow stresses than samples compressed without electrical current pulses.

Shear band induced fracture was observed for all compression samples. The corresponding fracture planes were inclined by approximately 45° towards the loading direction. Since the electrical current pulses were turned off before fracture occurred, it cannot be determined if fracture was initiated before or after the electric current pulses were switched off.

### 3.1.2 Temperature evolution during macroscopic electroplasticity experiments

Figure 3 shows the temperature evolution of sample 3 given in Figure 2 as a typical example for all electroplastically (EP) deformed samples. The temperature, which was measured at the sample surface in the centre of the gauge section, rises abruptly and then drops back to nearly room temperature after each electric current pulse.

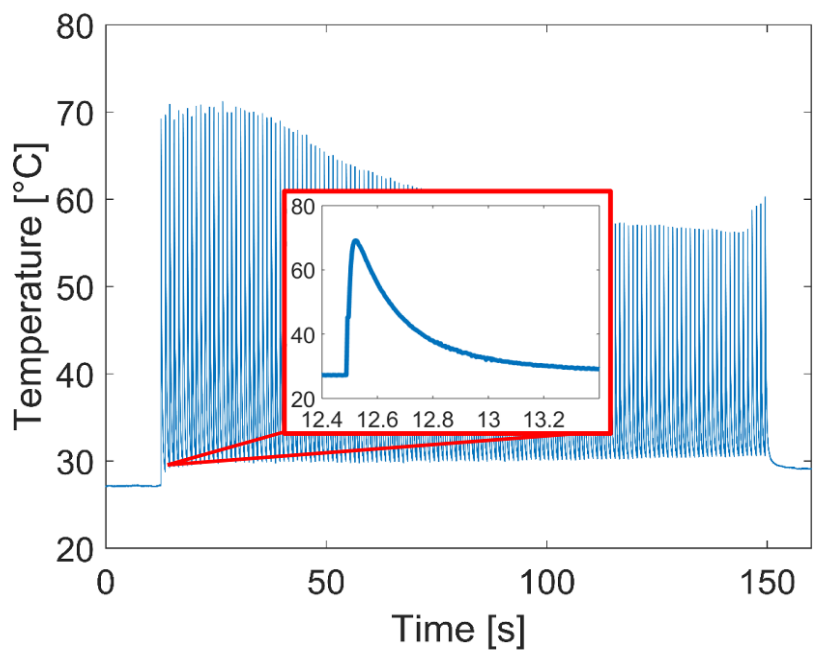

*Figure 3: Temperature evolution of sample 3 (cf. Figure 2) during electro-pulsed compression experiments.*

The maximum temperature measured during the EP compression tests was 76.0 °C (at a minimum temperature of about 29.5°C). The increase in minimum temperature measured during and after the EP compression experiments is assumed to be due to residual heat present in the sample.

### 3.2. Micromechanical electroplasticity experiments

### 3.2.1. Technical development

The development of an experimental setup enabling nanoindentation experiments with electric current pulses offers a fundamentally new approach to investigate the electroplastic effect. It allows us to evaluate the effect of electric current pulses on the mechanical properties and deformation mechanisms of defined microstructural constituents. Furthermore, due to the small dimensions it is possible to apply high current densities.

Figure 4 a) shows a schematic drawing of the commercial nanoindenter adapted for this setup (NanoTest P3, MicroMaterials Ltd.). The system is controlled by passing a current through the coil, which is then attracted towards the permanent magnet, resulting in a rotation of the pendulum around the frictionless pivot. This movement leads to the indenter tip being pushed towards/into the sample and can be tracked via the displacement sensor at the lower end of the pendulum. The new setup also contains a sample holder designed and built in-house, a dedicated tip holder developed for this application and a controllable power supply.

The newly designed sample holder ensures electric insulation of the sample towards the indenter due to the use of insulating ceramics and plastics (Figure 4 c)). The added stub for scanning electron microscopy (SEM) facilitates easy sample transfer to the microscope after indentation. The electrical parts of the tip holder are insulated towards the indenter by using an alumina rod (Figure 4 b)). A program was developed to start the current reproducibly at a desired time after the start of the indentation by automatically detecting the contact between tip and sample. Additionally, the achieved current values can be recorded and correlated with the mechanical data.

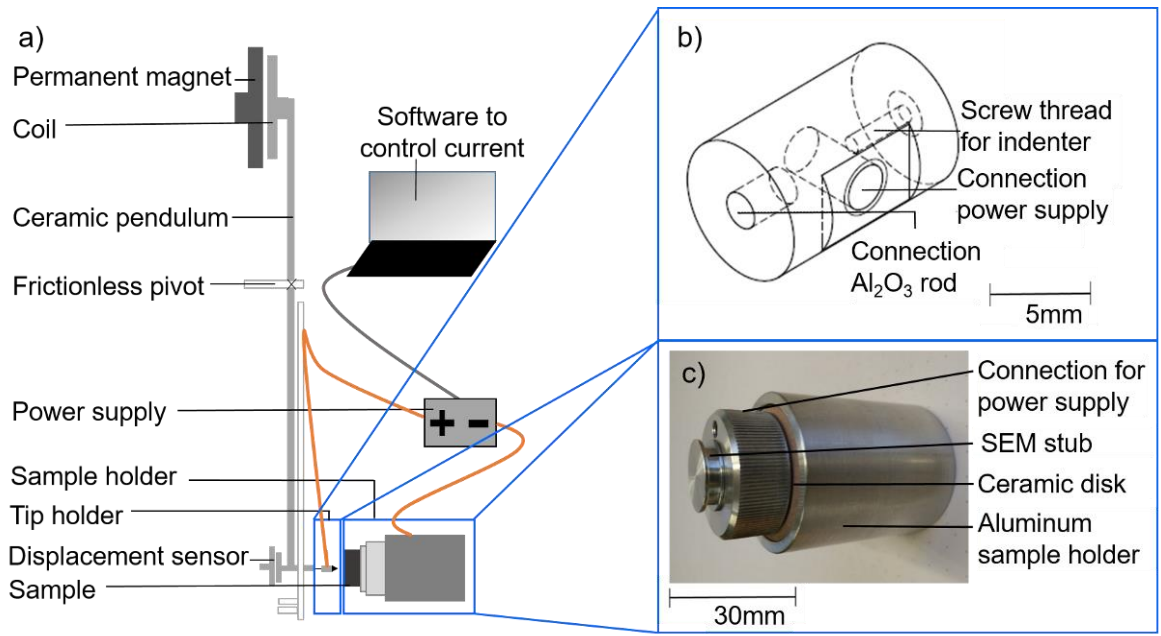

*Figure 4: a) Schematics of the new indenter setup built on a commercial nanoindentation platform, b) custom-built tip holder, c) custom-built sample holder.*

Similar to the macroscopic EP experiments, the electric current was applied in short pulses with a length of 100 µs each to minimize the effect of joule heating. It should be noted that the pulse duration applied in the micromechanical EP experiments is five times shorter than that applied during macroscopic electroplastic compression experiments to minimise joule heating at the increased current density. The current intensity was varied between 1.0 A and 3.0 A and applied at a defined frequency during loading and unloading (Figure 5). The indenter tip material was tungsten carbide, as it combines high electrical conductivity with high hardness.

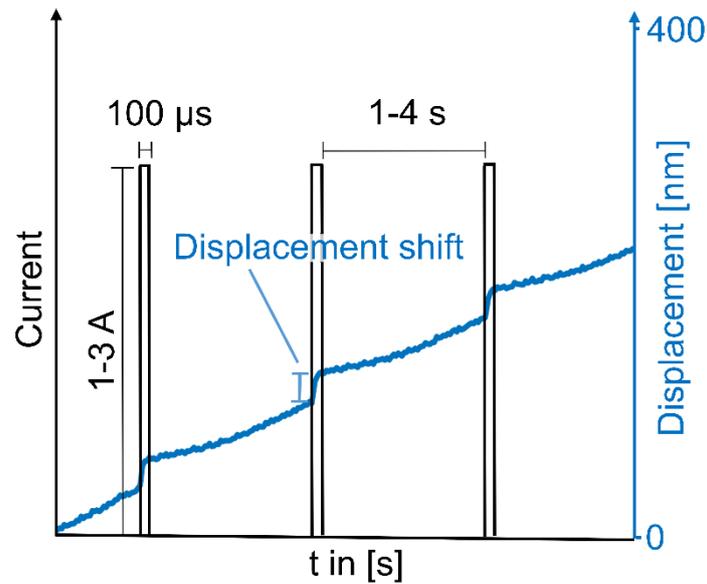

*Figure 5: Schematic of the applied current pulses as a function of time during the experiment with a measured displacement response on a eutectic Al-Al$_2$Cu sample. The experimental displacement data corresponds to a test conducted using a set point of 2 A per pulse.*

### 3.2.2. Nanoindentation experiments with different current intensities and pulsing intervals

The nanoindentation load-displacement curves obtained when applying different current intensities or pulsing intervals on a eutectic Al-Al$_2$Cu sample are shown in Figure 6. The application of electric current pulses during indentation causes a step-like increase in displacement with constant loading rate (see also Figure 5 above). A similarly step-like displacement curve is observed during unloading. During loading, these steps in the load-displacement curve are characterised by a rapid displacement shift towards higher displacements at constant load upon each electric current pulse, followed by a period of low displacement response to the applied load in the time between two consecutive electric current pulses. During unloading, the direction of the displacement shift is reversed and becomes more noticeable the smaller the remaining applied load (see section 3.2.3. Pulsed unloading experiments).

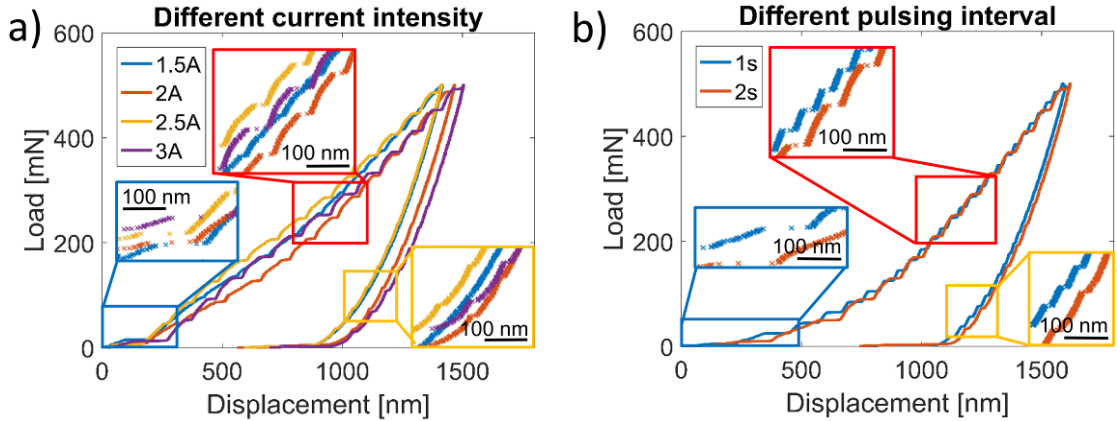

*Figure 6: Load-displacement curves on a eutectic Al-Al$_2$Cu sample with a) different current intensities of 1.5 A, 2 A, 2.5 A, 3 A at a loading and unloading rate of 20 mN/s, a pulse length of 100 µs and a pulsing interval of 2 s; b) different pulsing intervals of 1 s and 2 s, at a current intensity set to 1.5 A, a loading and unloading rate of 20 mN/s and a pulse length of 100 µs. The curves shown magnified are off-set by a load increment to improve discernibility.*

Variation of the current intensity reveals that the load-displacement curves corresponding to higher currents show "broader" steps, i.e. higher displacement shifts, than those corresponding to smaller currents (Figure 6 a)). As similarly observed during macroscopic EP compression experiments, the displacement shift induced by the first electric current pulse is larger than the displacement shifts induced by subsequent electric current pulses, independent of the applied current (blue box, Figure 6 a) and b)).

Figure 6 b) shows the effect of the interval between two electric current pulses on the load-displacement behaviour of the eutectic Al-Al$_2$Cu alloy. A larger pulsing interval results in larger displacement shifts but affects the maximum displacement values only marginally.

Indentations of the individual phases Al$_2$Cu-Θ and fcc-aluminium are shown in Figure 7 using also different current intensities and replicating the described effect that a greater current intensity yields larger displacement shifts. Due to the difference in hardness these experiments were conducted to different maximum loads to achieve comparable displacements. Similarly, the loading rates were scaled, using 20 mN/s on the Al$_2$Cu-Θ phase and 2 mN/s on the fcc-aluminium phase. Quantitatively, nanoindentation experiments on {110} ∥ ND (ND: normal direction) Al$_2$Cu-Θ phase grains reveal hardness values of 5.3 ± 0.09 GPa at a depth of 1100 nm for four indents performed without the application of electric currents and hardness values of 5.07 ± 0.53 GPa at a depth of 1100 nm for four indents performed with electric current pulses of 1.5 A ( a)). Note that for each value given here with a standard deviation, the individual data points are shown in Figure 8. As the number of indentations was limited to ensure tip wear would not be significant, the standard deviations are calculated only to convey a measure of repeatability or scatter and are not statistically relevant calculations with four indentations averaged on the Al₂Cu-Θ phase and two on the Al α-phase. The elastic modulus fitted for the upper 80% of the unloading curve ranged from 101.3 ± 3.6 GPa for indents without the application of electric current pulses to 118.5 ± 2.9 GPa for indents with 1.5 A current pulses during application (Figure 8 b)). The hardness values for the Al α-phase (oriented $\{3\bar{5}2\}$ ∥ ND) accounted for 0.34 ± 0.00 GPa for two indents performed without current pulses and 0.33 ± 0.01 GPa for two indents performed with 1.5 A current pulses during deformation (Figure 8 c)). The elastic modulus amounted to 72.6 ± 1.4 GPa without application of current pulses and 70.4 ± 1.3 GPa for indents performed with 1.5 A (Figure 8 d)).

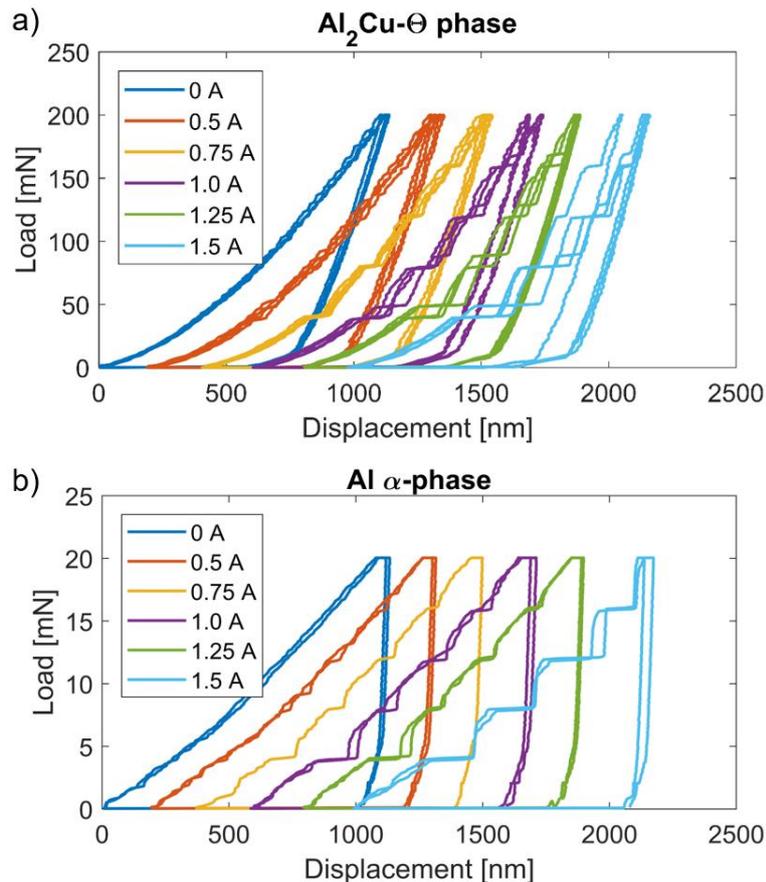

Figure 7: Load-displacement curves of the a) Al₂Cu-θ and b) Al α-phase with different applied current intensities during loading with a loading rate of 20 mN/s for (a)) and 2mN/s b)), a pulse length of 100 µs and a pulsing interval of 2 s. The curves were offset by 200 nm, respectively, to ease differentiation.

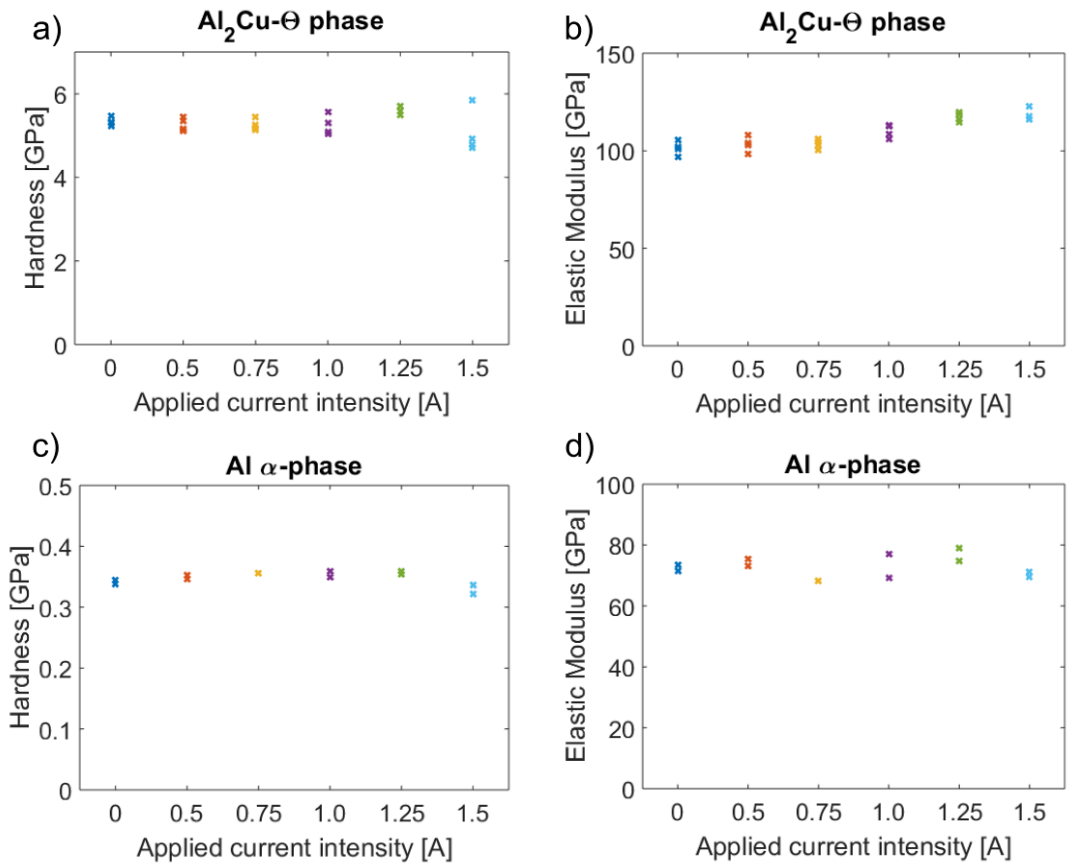

*Figure 8: Development of the hardness and elastic modulus over applied current intensity for the five indents in Al₂Cu-θ (a), b)) and two indents in α-Al (c),d))-phases (note: for 0.75 A: current pulses were only applied during one indent).*

<u>3.2.3. Pulsed unloading experiments</u>

The application of electric current pulses of 2.5 A during unloading causes a similar step-like displacement response as observed during loading. Figure 9 shows load-displacement curves with loading and unloading rates of 10 mN/s and 20 mN/s, respectively. The displacement shifts occurring during loading and unloading at a loading (and unloading) rate of 10 mN/s are smaller and half in height compared the displacement shifts occurring at a loading (and unloading) rate of 20 mN/s. The direction of the displacement shifts changes during unloading, specifically, the first displacement shift during unloading continues to higher displacement (blue box, Figure 9), but changes towards lower displacement during progressive unloading (red box, Figure 9). Furthermore, the displacement shifts during unloading are smaller than those during loading and increase as the applied load decreases. The maximum displacement seems to be unaffected by the loading rate.

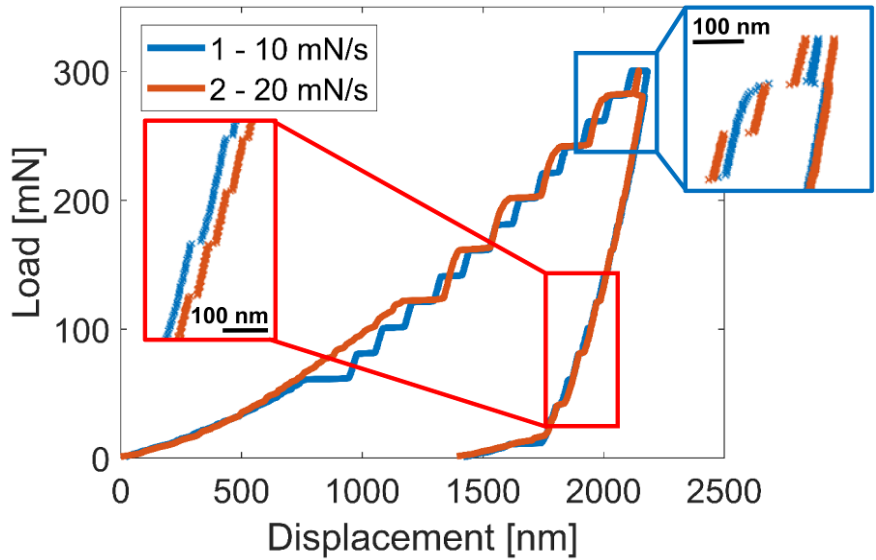

*Figure 9: Load-displacement curves with different loading and unloading rates of 10 mN/s and 20 mN/s, respectively. The current intensity was set to 2.5 A and the pulse length to 100 µs. The red and blue boxes show displacement steps during unloading. The curves in the red box are offset by a load increment to enhance distinctiveness.*

## 4. Discussion

### 4.1. Macroscopic EP experiments

Our experimental results reveal that electric current pulses can reduce the stresses needed for plastic deformation and enhance plastic deformation of a eutectic Al-Al$_2$Cu alloy under compressive load. The stress-strain curves of the EP deformed specimens show rapid stress drops upon each electric current pulse. These stress drops are followed by a sudden increase in stress until the next electric current pulse was applied. A similar phenomenon has been reported by several other researchers before [1, 2, 34, 35] and confirm the applicability of EP assisted forming for Al-Cu eutectic alloys.

The EP stress-strain curves further reveal that the stress drop induced by the first electric current pulse amounts to more than two times the stress drops induced by the following electric current pulses. A similar effect has been reported by Livesay et al. [36], who investigated the effects of electrical current pulses on the plastic deformation of Al thin films. Their results [36] reveal that the increase in elongation induced by the first current pulse is higher than those induced by subsequent electric current pulses.

The high stress drop induced by the first electric current pulse observed in the current study might be caused by the electron wind effect [1, 5, 11, 12], which is assumed to help dislocations to overcome obstacles. We therefore assume that during the first current pulse dislocations present in the microstructure are de-pinned from short-range obstacles, hence, inducing a major stress drop. During the following electric current pulses, the number density of pinned dislocations is consequently lower, resulting in smaller stress drops during subsequent electric pulsing. The slope of the EP compressed samples is the same as the slope of samples compressed without electric current pulses indicating that short electric current pulses (0.5 ms) have only a short-term effect and do not influence the overall strain hardening behaviour of the eutectic Al-Cu alloy investigated. However, for a continuous electric field, Conrad et al. [5] observed a lowered strain hardening in metals. Similarly, Liu et al. [37] have reported a decrease of the strain hardening exponent with increasing current density for a TRIP steel when applying electric current pulses during deformation. However, in our experiments, the interaction of electrons with dislocations and interfaces are expected to dominate the EP deformation behaviour, whereas in the latter, the TRIP effect might show additional and / or different EP-induced mechanisms.

The observed maximum temperature induced by electric current pulses amounts to 76°C and after each applied electric current pulse the temperature decreased again to near room temperature. Since the heat loss due to radiation from the sample surface can be neglected [38], it is assumed that the measured surface temperature corresponds to the temperature inside of the sample.

This temperature increase causes thermal expansion of the samples, which can be estimated by applying the rule of mixture for the thermal expansion coefficient of the composite material giving the composite coefficient as $\alpha_C = 2.38 \times 10^{-5}$ 1/K with the thermal expansion coefficients of the constituents taken as $\alpha_{Al} = 2 \times 10^{-5}$ 1/K and $\alpha_\theta = 2.7 \times 10^{-5}$ 1/K [39] and volume fractions of $X_{Al} = 0.55$ and $X_\theta = 0.45$ taking into account the area fractions determined from light microscopy images. The induced thermal strain, $\varepsilon_{th}$, was then calculated according to Eq. 1:

$$\varepsilon_{th} = \alpha_C \times \Delta T \qquad\qquad \text{Eq. 1}$$

assuming a maximum temperature increase, $\Delta T$, of 46.5 °C. The resulting maximum thermal strain amounts to 0.11%. The modulus of the composite, $E_C$, was also calculated using the rule of mixtures with $E_{Al}$ = 75.15 GPa and $E_\theta$ = 99.28 GPa [23] giving $E_C$ = 86.01 GPa. The additional thermal stress, $\sigma_{th}$ was calculated using Eq. 2

$$\sigma_{th} = E_C \cdot \varepsilon_{th} \qquad \qquad \text{Eq. 2}$$

to be 94.5 MPa.

When considering these induced thermal expansion and thermal stresses, the change in stress upon electric current pulsing is assumed to be due to three different effects, namely, the thermal expansion of the sample, the thermal softening of the material and the electroplastic effect. These results were not corrected for the unknown thermal expansion of the machine.

Thermal expansion of the samples induces an increase in stress, whereas thermal softening and the EPE cause a stress reduction. The stress increase due to thermal expansion amounts to 94.5 MPa at the maximum temperature observed during EP compression experiments when assuming that the temperature measured at the surface of the sample is similar to the temperature inside the sample.

In order to quantify the effect of thermal softening during the experiments, hardness measurements of Al and $Al_2Cu$ at RT and 200°C reported by Chen et al. [40], were further analysed. The hardness of the intermetallic $Al_2Cu$ phase decreased from 5.77 GPa at room temperature to a value of 5.33 GPa at 200°C whereas the hardness of the aluminium phase decreased from 1.45 GPa at RT to 0.68 GPa at 200°C [40]. When assuming that the deformation mechanisms for both materials are unaffected by a temperature increase to 200°C, the hardness drop can be calculated via linear interpolation. The assumption that the active deformation mechanisms are not affected by an increase in temperature to 200°C is based on the results by Ball et al. [41] for Al and by Chen et al. [40] for $Al_2Cu$ who reported no changes in hardness within the standard deviation indicating no change in the active deformation mechanisms. The resulting hardness value of $Al_2Cu$ at 76°C is accordingly calculated to amount to 5.66 GPa and the one of Al to amount to 1.25 GPa. The resulting composite hardness amounts to 3.24 GPa according to the rule of mixture. This corresponds to a decrease in hardness of 0.16 GPa when the temperature increases from room temperature to 76°C. When assuming a small strain hardening, this loss in hardness, H, can be converted into a decrease in yield strength, $\sigma_y$, according to Eq. 4 [42]:

$$H = 3 * \sigma_y \qquad\qquad \text{Eq. 4}$$

It should be noted that the assumption of a low strain hardening coefficient overestimates the yield strength. The resulting upper bound on the decrease in yield stress due to thermal softening amounts to 52.8MPa. As has been mentioned above, the stress increase induced by thermal expansion amounts to 94.5 MPa.

The stress drops measured during EP assisted compression experiments amount to 47.8 ± 4.8 MPa for the first electric current pulse and to 18.3 ± 1.6 MPa for the subsequent electric current pulses. When considering a decrease in stress due to thermal softening of 52.8 MPa and an increase in stress due to thermal expansion of 94.5 MPa, the softening induced by the EPE amounts to about 89 MPa during the first electric current pulse and about 60 MPa during subsequent electric current pulses. This supports the hypothesis that the observed stress drops are not solely induced by thermal softening but that the EPE causes an additional stress decrease. This assumption is further supported by the fact that the temperature increase observed during the first electric current pulse is similarly high as those observed during the subsequent electric current pulses. Therefore, the large stress drop occurring during the first electric current pulse cannot be a result of merely thermal softening.

## 4.2. Micromechanical EP experiments

### 4.2.1. Nanoindentation experiments with different current intensities and different pulsing intervals

An increase in the applied current intensity (Figure 6 a) for the eutectic microstructure and Figure 7 for the individual phases) results in larger displacement shifts at constant load. This observation is assumed to be caused by enhanced plasticity induced by electric current pulses.

Similarly, Okazaki et al. [10] and Troitskii [43], have reported larger stress drops with increasing current density. In the present study, an increase in current intensity is interpreted as increased current density at a constant pulse length.

The observed periods of low displacement response, which follow the rapid displacement shifts upon electric current pulsing are interpreted as the response of the material after the effect of the short-term electric current pulse is exhausted.

The observation that the applied current intensity influences the magnitude of the displacement shift indicates a dislocation motion related EPE mechanism. This is assumed to be either related to the de-pinning of dislocations from obstacles or an increased dislocation mobility or a combination of both.

Independent of the applied current intensity, the displacement shift induced by the first electric current pulse during indentation is significantly larger than those of the subsequent electric current pulses. This is consistent with our macroscopic EP assisted compression experiments that show that the first electric current pulse induces a stress drop that is more than two times larger than the subsequent electric current pulses. As has been discussed above, this effect is assumed to be most likely related to the de-pinning of dislocations from short-range obstacles induced by electric current pulses.

A longer inter-pulse interval results in larger displacement shifts (Figure 6 b)). This further supports our hypothesis that electric current pulses might induce the de-pinning of dislocations from short-range obstacles: a longer inter-pulse interval corresponds to a longer time between two electric current pulses during which more dislocations might become pinned than during shorter inter-pulse intervals. Consequently, more dislocations are assumed to be de-pinned during a subsequent electric current pulse resulting in a larger displacement shift.

Similarly, Troitskii [6] has reported that increasing the inter-pulse interval resulted in higher stress drops up to an interval of 5 seconds, beyond which no further increase in the stress drop magnitude for higher inter-pulse intervals was observed.

The measured hardness of the $Al_2Cu$-Θ phase at a depth of 1100 nm amounts to $5.3 \pm 0.09$ GPa without application of current pulses and $5.07 \pm 0.53$ GPa with application of 1.5A (Figure 8 a)). The hardness value without application of current pulses corresponds well to those observed by Chen et al. [40] who reported hardness values of about 5.77 GPa at room temperature. The small differences might arise from the different solid solution content of the alloys as well as different crystallographic orientations. The observed decrease in hardness upon application of electric current pulses and increased standard deviation is assumed to be caused by experimental scatter due to the application of electric current pulses. Specifically, a current pulse might affect the measured hardness due to displacement shift and at the time increase the standard deviation (see Figure 7). The elastic moduli measured for the $Al_2Cu-\Theta$ phase without application of electric current pulses correspond well to the results by Chen et al [40]. With applied electric pulsing, no change in elastic modulus is expected for drift electron – dislocation interactions [44] as these do not influence the atomic bonding strength that determines the material's elastic response and therefore elastic modulus value.

However, a small increase in average modulus was observed for the $Al_2Cu-\Theta$ phase. A closer inspection of the load-displacement curves (Figure 7a)) revealed that a different and linear gradient is indeed observed towards the top of the unloading curve. This indicates that the increase in modulus is not related to any time-dependent relaxation processes after the final pulse, as seen regularly in indentations affected by creep. Whether this unexpected increase in stiffness is related to the electroplastic pulsing, e.g. through changes in the pile-up behaviour, or artefacts, e.g. tip wear, could not yet be resolved. However, we assume that the electroplastic effect is not influencing the atomic bonding strength, as the effect is mostly related to the interaction of the electric current with dislocations [44].

The hardness values obtained for the fcc-Al phase amounted to 0.34 ± 0.00 GPa (Figure 8 c)), which is consistent with hardness values reported in literature where Liu et al. [45] reported hardness values of 0.47 GPa at an indentation depth of 100 nm for Al and Pharr et al. [46] reported a hardness value of 0.21 GPa for Al. The small deviations in the hardness values reported in the literature are assumed to arise from the different crystal orientations as well as different compositions investigated.

The elastic modulus measured for the Al α-phase remained approximately constant, as expected, and the value of 72.6 ± 1.4 GPa without application of current pulses corresponds well to literature where modulus values of 68.0 GPa [47] and 70.4 GPa [46] are reported.

The magnitude of the displacement shifts induced by electric current pulses during nanoindentation is observed to be rate-dependent (Figure 9): a high loading rate induces larger displacement shifts during loading and unloading than a low loading rate when the same current intensity is applied.

This observed strain rate dependence cannot be explained by merely thermally activated dislocation motion. Higher loading rates would reduce the probability of dislocations to overcome the Peierls barrier by thermal fluctuations [48], hence, resulting in increased yield and flow stresses rather than decreased flow stresses as observed in the present study. This again indicates that the observed EPE cannot be only induced by thermal activation.

Further, the strain rate dependence of the EPE is controversially discussed in the literature [44-48]. Specifically, Troitskii [49], Ross et al. [50] and Varma et al. [51] have reported that higher current densities are required to induce the same EPE at higher strain rates indicating that the EPE decreases with increasing strain rate. In contrast, Cao et al. [52] have observed an increasing EPE with increasing strain rate for Nb, whereas Okazaki et al. [53] have observed no strain rate dependence of the EPE. It is therefore assumed that the strain rate dependence of the EPE is not controlled by a general mechanism but is controlled by the material and the predominant deformation mechanisms.

### 4.2.2. EP unloading experiments

Our nanoindentation experiments further reveal that electric current pulses during unloading also induce load-displacement steps, Figure 9 and Figure 10. The direction of the displacement shifts changes from displacement shifts towards higher displacements to displacement shifts towards lower displacements during unloading. Thermal effects such as thermal expansion of the tip and / or sample would not induce a change in the direction of the displacement shifts. It is therefore assumed that the displacement shifts during unloading are caused by increased reverse plasticity generated by the electric current pulses. A similar observation has been reported by Okazaki et al. [54] for polycrystalline titanium wires (Figure 10 a)). They [54] observed short-term load drops in the load-time curve upon electric current pulsing during unloading in a tensile test. In the beginning of unloading, an electric current pulse induced a short-term load drop followed by an increase in load to a level slightly below the load level before the electric current pulse, $\sigma_{r1}$. Whereas in the end of unloading, the electric current pulses induced load drops which were followed by an increase in load to load levels above the load before the electric current pulse was applied, $\sigma_{r3}$. They [54] attributed the latter to reversed plastic flow.

This agrees to our observation that the direction of displacement shifts induced by electric current pulses changes its direction from towards higher displacements to towards lower displacements, Figure 10.

Okazaki et al. [54] suggested that internal long-range stresses in the material are responsible for this effect. Specifically, if unloading exceeds the level of internal long-range stresses, electric current pulsing induces forward plastic flow, whereas at load levels below the existing internal long-range stresses reversed plastic flow is induced by an electric current pulse [54].

In the eutectic composite microstructure investigated in the present study, the two phases deform not homogeneously and strain gradients are assumed to form during deformation, particularly in the vicinity of phase boundaries. These strains and those induced upon casting due to different thermal expansion coefficients of the constituent phases existing in the eutectic microstructure might induce long-range internal back stresses on dislocations [39, 55]. Hence, electric current pulses might promote the motion of dislocations under these long-range internal stresses and cause the observed displacement shifts during unloading. Thermal softening alone would not cause such back stresses during unloading.

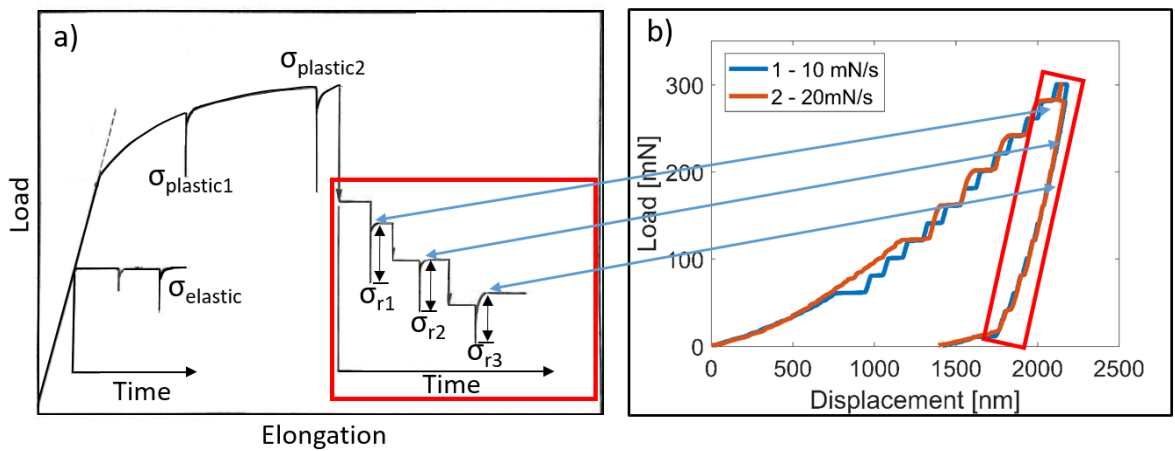

*Figure 10: Comparison of the EPE during unloading of a tensile test and unloading during nanoindentation, (a) adapted from Reference [54] with permission by Elsevier, a) Load-elongation curve of a tensile test on a polycrystalline titanium wire, current pulses ($12*10^3$ A/mm$^2$, 80 µm) were applied during the test; b) nanoindentation load-displacement curve on a eutectic Al-Al$_2$Cu sample, electric current pulses (3 A, 100 µs) were applied during loading and unloading.*

The difference in the maximum displacement when compared to the experiments shown in Figure 6 and Figure 9 are assumed to be caused by the composite microstructure, where the phase fractions and crystallographic orientations present in the microstructure below the indents vary.

# 5. Conclusions

We have performed macroscopic and micromechanical electroplastic deformation experiments on a eutectic Al-Al$_2$Cu alloy. To this end, we have developed a new nanoindentation setup to apply electric current pulses during nanoindentation. From our study we draw the following conclusions:

- Macroscopic compression experiments using electric current pulses (current density 1.66 ± 0.01 kA/mm$^2$, pulse length 0.5 ms) show stress drops upon electric current pulses, as well as higher fracture strains and lower flow stresses than during compression experiments without electric current pulses.

- The temperature increased from room temperature to up to 76°C due to electric current pulses during macroscopic EP compression tests.

- Both, macroscopic and nanoindentation EP experiments show that the first current pulse induces a larger stress drop than the subsequent electric current pulses. This is assumed to be caused by the de-pinning of dislocations from obstacles.

- In electroplastic nanoindentation experiments (pulse length 100 µs), a higher current intensity, pulsing interval and loading rate result in larger displacement shifts. This indicates an increased dislocation mobility and de-pinning of dislocations from obstacles.

- Electric current pulses during unloading induce displacement shifts that change their direction during unloading from towards higher displacements to towards lower displacement. This is assumed to be caused by long-range internal stress fields present in the deformed microstructure.

# Acknowledgement

The authors gratefully acknowledge funding of the priority program "Manipulation of matter controlled by electric and magnetic field: Towards novel synthesis and processing routes of inorganic materials" (SPP 1959/1) by the German Research

Foundation (DFG). This work was supported by grant number 319419837 and grant number 319282412.

## Data Availability Statement:

The raw data required to reproduce these findings are available on request from the corresponding author of this study.

# Investigation of the electroplastic effect using nanoindentation

D. Andre[1], T. Burlet[1], F. Körkemeyer[2], G. Gerstein[2], S. Sandlöbes-Haut[1], S. Korte-Kerzel[1]*

[1]Institute of Physical Metallurgy and Metal Physics, RWTH Aachen University,

[2]Institute of Materials Science, Leibniz University Hannover

\* korte-kerzel@imm.rwth-aachen.de (corresponding author)

## Supplementary materials

### Adjustment of current densities

With increasing indentation depth, the contact area between tip and indented material increases. To keep the current density constant, the current intensity has to be increased as well. When assuming that the size of a load / stress drop (corresponding to hardness drops during nanoindentation) depends only on the applied current density [1-4], the relevant contact area in nanoindentation must be evaluated, as this is not necessarily the same as considered in the mechanical analysis for Berkovich indentation, i.e. A=24.5*h² with h being the penetration depth.

We have conducted experiments to control the current density during nanoindentation. Specifically, to maintain a constant current density with increasing load and displacement, we considered two different ways for the calculation of the current density based on (i) the surface area of the employed Berkovich indenter tip and (ii) the surface area of the plastic zone. The Berkovich surface considers the direct contact area between tip and sample (Figure S1 a)) and scales according to

$$A_B = 26.96 * h^2 \qquad\qquad \text{Eq. 5.}$$

The radius of the hemispherical area surrounding the plastic zone is assumed to scale with a factor of 2.2 with the contact radius [5] which results in a surface function that scales approximately with (Figure S1 b)),

$$A_P = 116.03*h^2 \qquad\qquad Eq.\ 6.$$

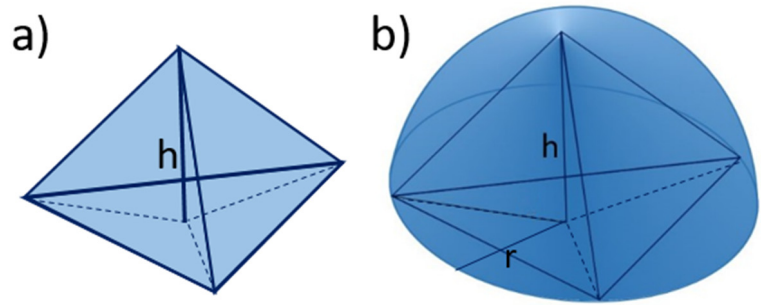

*Figure S1: Considered approaches to calculate the current density during nanoindentation: a) Berkovich surface function: $A_B = 26.96*h^2$, b) plastic zone function: $A_P = 116.03*h^2$*

Figure S2 shows the current-displacement (a) and corresponding load-displacement (b) curves with automatically adjusted current intensity of the electric current pulses to a constant current density of 10 kA/mm$^2$ during nanoindentation using the Berkovich function. However, as evident from Figure S2 a), at small displacements the current intensities are higher than the overall trend. This might be due to short pulse intervals making it difficult to exactly control the current intensity. The corresponding load-displacement curve (Figure S2 b)) shows load drops upon electric current pulses. A threshold value was set to detect load drops and, consequently, anomalies in the load-displacement curves, therefore, small load drops were not detected. Consequently, the number of applied current pulses does not correspond exactly to the number of detected load drops. Similar results were obtained when using the plastic zone function.

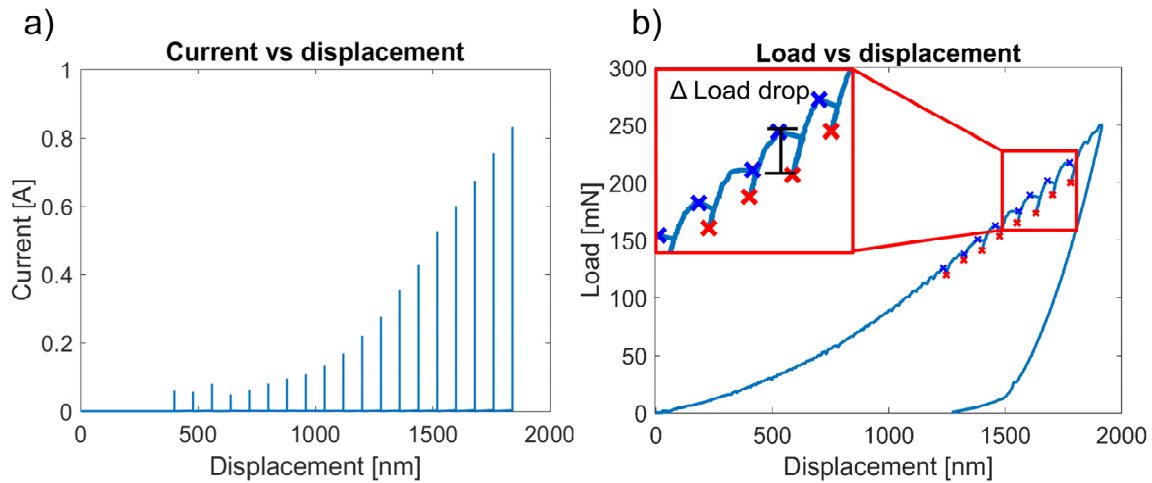

*Figure S2: a) Current-displacement curve obtained during EP indentation, the current intensity of each pulse was calculated to achieve a constant current density of 10 kA/mm² using the Berkovich surface function. b) Corresponding load-displacement curve revealing load drops upon electric current pulses. The displacement rate was set to 20 nm/s, the pulse length to 100 µs, and the pulsing interval to 4 s. Blue and red marks show automatically detected maximum and minimum load drop values and red boxes indicate the magnitude of a load drop.*

Figure S3 compares the applied current intensities and hardness drops achieved during nanoindentation when applying the Berkovich function (Figure S3 a)) and the plastic zone function (Figure S3 b)) to adjust the current intensities to a constant current density of 10 kA/mm². For the Berkovich function 11 indents and for the plastic zone function 30 indents were considered, the higher number of indents used for the plastic zone function is due to the high scatter present in the data, Figure S3 b). The maxima of the current intensity were fitted with a 2nd order polynomial function and the hardness drops were fitted using a linear function since this allows to compare the slopes of both functions. The corresponding fitting functions are given in the graphs. The standard deviation of the current intensity maxima amounts to 0.062 A for the Berkovich function and to 0.119 A for the plastic zone function. The larger standard deviation of the plastic zone function is consequently related to a larger standard deviation of the hardness drops of 0.271 GPa for the plastic zone function compared to 0.045 GPa for the Berkovich function.

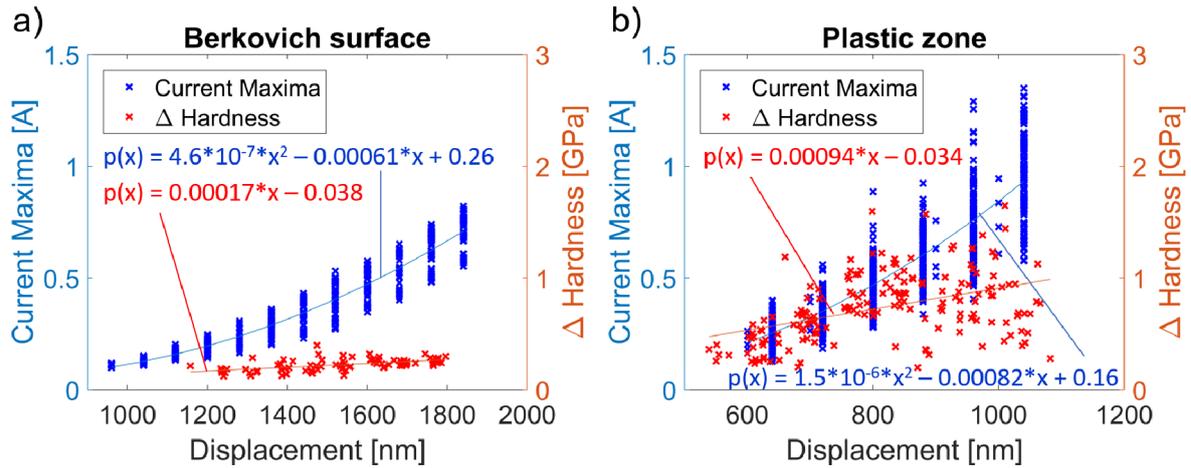

*Figure S3: Applied current intensities and resulting hardness drops over displacement for current intensity increase according to the plastic zone function (30 indents were considered) and according to the Berkovich function (11 indents were considered)*

In the present study, we observe that the slope of the resulting hardness drops is smaller for the Berkovich surface function (Figure S3) than for the plastic zone function. This might indicate that the Berkovich surface function is a better choice for calculating the relevant contact area for electroplasticity experiments during nanoindentation. However, the difference in slope is small and a large scatter of the applied current intensities was measured. The latter is due to the fast increase in current intensity during the test. It manifests also in a large experimental scatter, which is increased further by the different intrinsic properties of the Al and the $Al_2Cu$ phases in the eutectic affecting the size of the observed load drops.

We therefore further evaluated the drops in hardness induced by different current intensities in the $Al_2Cu$-θ phase (see Chapter 3.2.2, Figure 7 a)). For this evaluation, the first hardness drop was neglected due to the observation that the first current pulse induces a larger drop in hardness (see Chapter 4.2.1). All following hardness drops, together with the applied current intensities and contact depths at time of pulse application enabled us to account for the current density when considering both, the Berkovich surface function and the plastic zone function. The resulting hardness drops are compared for different current density intervals (Figure S4).

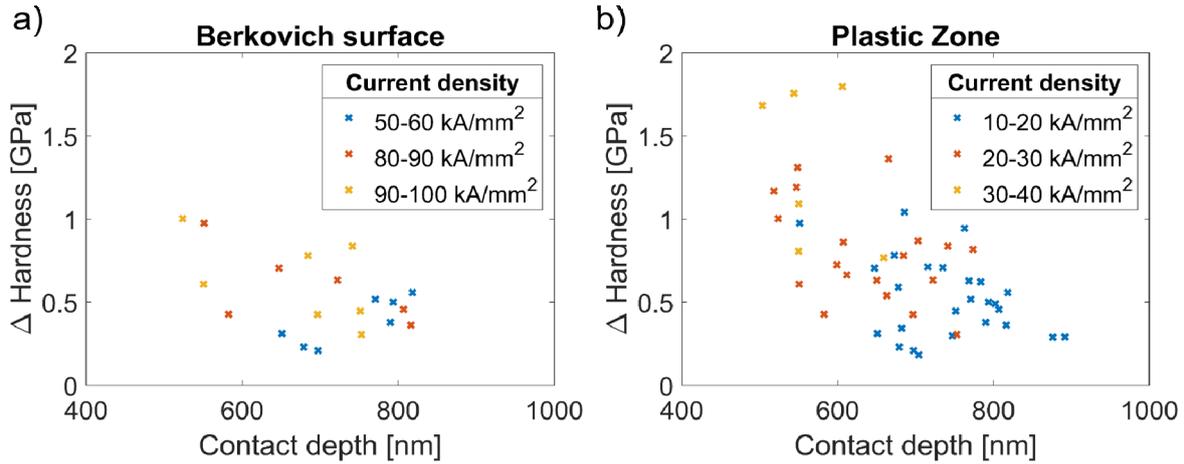

*Figure S4: Hardness drops induced by electric current pulses in the Al₂Cu-θ phase (for load-displacement data, see Figure 7a)) over contact depth. a) hardness drops induced by current pulses for current density intervals calculated using the Berkovich surface function, b) hardness drops induced by current pulses for current density intervals calculated using the plastic zone surface function.*

However, neither the hardness drops resulting from current pulses of one current density interval for the Berkovich surface function (Figure S4 a)), nor for the plastic zone function (Figure S4 a)), remain constant. Therefore, we cannot explicitly conclude which function is the relevant surface function for conducting electroplasticity nanoindentation experiments.

One reason for this in terms of plasticity mechanisms might be the change in strain rate during indentation, which also has an impact on the size of the EPE (see Chapter 3.2.3) and is known to affect hardness in thermally activated processes such as grain or phase boundary deformation (e.g. in the eutectic) or Peierls potential controlled flow (e.g. in the Al₂Cu-θ phase).

Furthermore, it is still controversially discussed how load / stress drops during EP forming are related to the current density. In addition to the assumption that the load drop magnitude is linearly dependant on the current density [1-4], Molotskii [6] has proposed that the stress drops scale with the square of the current density. Further, Liu et al. [7] have shown that the actual strain might influence the magnitude of stress drops during EP deformation as they observed that the flow stress reduction increases with increasing strain in advanced high strength steels. They [7] have assumed that this strain dependence is related to the larger density of dislocations in the microstructure resulting in a more pronounced EPE. In contrast, Conrad et al. [8] reported a decreased EPE with higher applied strains for copper and aluminium. Conrad et al. [8], [9] and Sprecher et al. [10] suggested a direct proportionality between the electron wind force and the current density, but an additional square proportionality of another, yet unclear, effect. However, these and additional thermal effects were not considered in the contact area validation. With respect to thermal effects, we consider the induced short-term temperature peaks being not significant in nanoindentation experiments due to the high thermal conductivity of both constituents and the large amount of bulk material around the indent. Due to the obviously complex and not yet well-understood relation between the current density (and possibly other parameters) with the magnitude of stress drops, verification of a relevant function to calculate a constant current density for varying indentation depth in nanoindentation is not possible to date.